# Observing Molecular Hydrogen Clouds and Dark Massive Objects in Galactic Halos

F. De Paolis[1,2,*], G. Ingrosso[1,2,*], Ph. Jetzer[3,4,**], A. Qadir[1,5], and M. Roncadelli[6]

[1] Dipartimento di Fisica, Università di Lecce, Via Arnesano, CP 193, 73100 Lecce, Italy

[2] INFN, Sezione di Lecce, Via Arnesano, CP 193, 73100 Lecce, Italy

[3] Institute of Theoretical Physics, University of Zurich, Winterthurerstrasse 190, CH-8057 Zurich, Switzerland

[4] Institute for Nuclear and Particle Astrophysics, LBL and Center for Particle Astrophysics, University of California, Berkeley, USA

[5] Department of Mathematics, Quaid-i-Azam University, Islamabad, Pakistan

[6] INFN, Sezione di Pavia, Via Bassi 6, I-27100, Pavia

January 20, 1995

**Abstract.** Molecular hydrogen clouds can contribute substantially to the galactic halo dark matter and may lead to the birth of Massive Halo Objects (MHOs) observed indirectly by microlensing. We present a method to detect these molecular clouds in the halo of M31 using the Doppler shift effect. We also consider the possibility to directly observe MHOs in the halo of M31 via their infrared emission.

Send offprint requests to: Ph. Jetzer

*This work has been partially supported by Italian Space Agency

**Supported by the Swiss National Science Foundation

## 1. Introduction

Observations of microlensing events (Alcock et al. 1993, Aubourg et al. 1993) towards the Large Magellanic Cloud (LMC) support the idea that Massive Halo Objects (MHOs) may provide a substantial part of the halo dark matter (De Rújula et al. 1992 and references therein). In effect, current data on microlensing do not yet allow to decide whether MHOs are dark matter constituents of the galactic halo or faint stars in the galactic disk or in the LMC itself (De Rújula et al. 1994 and Gates et al. 1994). Nevertheless, recent observations of the Hubble Space Telescope have ruled out the possibility that faint red stars can explain all the microlensing events, while these observations do not rule out smaller halo objects like brown dwarfs (Bahcall et al. 1994).

Recently, a scenario (De Paolis et al. 1994) has been proposed which accounts for the formation of MHOs in the galactic halo. Moreover, a significant fraction, $f$, of the initial gas is expected to be present today in the halo (as the conversion efficiency of the primordial gas into MHOs could scarcely have been 100%). It has been argued that this gas should be in the form of molecular clouds, which are notoriously difficult to detect (Pfenniger et al. 1994). The aim of this paper is to discuss how these molecular clouds as well as MHOs can be observed directly in the nearby M31 galaxy.

We begin by addressing the issue of detection of molecular clouds. Our picture (De Paolis et al. 1994) envisages dark clusters of mass $\sim 10^6\ M_\odot$ consisting of MHOs and molecular clouds of mass $\leq 1\ M_\odot$. These clusters are expected to be present at a distance of $10 - 20$ kpc up to $\sim 100$ kpc with mean density distribution $\rho(r) \sim r^{-2}$. The number of molecular clouds per cluster ought to exceed $10^6\ f$. It has been pointed out that a signature of this scenario is the $\gamma$-production by cosmic rays striking the molecular clouds (De Paolis et al. 1994, Luo and Silk 1994). However, since the knowledge of the cosmic ray intensity and energies in the galactic halo is quite poor, alternative detection methods are welcome.

## 2. Alternative detection methods for molecular clouds in M31

In the past, molecular clouds were hotter than the CBR. How far from the CBR temperature are they now? An upper limit can be set by considering the anisotropy they would introduce in the CBR due to their higher temperature, assuming that the clouds are optically thick at every frequency. Looking at a region of the sky away from the galactic centre and off the galactic disc, we expect to see about a dozen of dark clusters per square degree. Here we suppose, for illustration, that dark clusters are spherically distributed at of mass $m$) and radius (from the virial theorem) $r \sim 2 \times 10^{-1} (m/M_\odot)$ pc, corresponding to an angular size of $\sim 1.8'$. Accordingly, due to the low surface filling factor $S \sim 10^{-2}$ ($S$ is the ratio of filled to total surface in a field of view), the ratio of the temperature excess of the clouds to the CBR temperature cannot be more than $\sim 10^{-3}$. This bound can further be strengthened through observational limits on the anisotropy of the CBR on the angular scale of $\sim 10'$ (Toffolatti et al. 1994), due to the corresponding increase of $S$.

Realistically, molecular clouds are not optically thick at every frequency and therefore they cannot be regarded as a black body. Indeed, a molecular cloud close to the CBR temperature emits a set of lines due to molecular rotational transitions. If we consider clouds with cosmological primordial composition, the only molecule that contributes to the microwave band with optically thick lines [1] is LiH (Maoli, Melchiorri & Tosti 1994). However, the precise chemical composition of the molecular clouds in the galactic halo is unknown. Since, up to now, no material with zero metal abundance has been found at low redshift in the Universe (see, e.g., Kunth et al. 1994), molecular clouds in dark clusters would contain heavy molecules made of metals formed during a first chaotic galactic phase. Even if the heavy molecule abundance is very low (as compared with the abundance in interstellar clouds), the lines corresponding to the lowest rotational transitions are optically thick [2].

Consider now molecular clouds in M31, for which we assume the same scenario outlined above for our galaxy. Our distance from M31 is $D \sim 650$ kpc. Although it is not clear what is the rotational velocity of the M31 halo (and also the radial anisotropy degree

---

[1] Taking a cosmological abundance with $n_{LiH}/n_{H_2} \sim 10^{-10}$ and the resonance cross-section $\sigma_\nu \sim 5 \times 10^{-10}$ cm$^2$ at $T = 2.76$ K, the optical depth $\tau_\nu = \int \sigma_\nu \, n_{LiH} \, dl$ (the integral being performed along the line of sight intercepted by molecular clouds) corresponding to the lowest-lying rotational transition of LiH at the frequency $\nu_0 = 444$ GHz implies that this line is optically thick. This result has been verified for a wide range of molecular cloud masses and densities.

[2] If we assume a metal abundance of $\sim 10^{-3}$ with respect to that in the $\zeta$ Ophiuchi cloud (Viala et al. 1987) we find that the CO molecule, for example, has a set of optically thick lines in the microwave band. The same could well happen for other molecules, but the calculation of the exact emission spectrum of molecular clouds (which must depend also on the incoming background radiation field) is beyond the scope of the present work.

of 50 − 100 km s$^{-1}$.

As molecular clouds are at a temperature close to that of the CBR, they would be indistinguishable from the background. But - since the clouds are moving - there is a Doppler shift. At worst, we expect at least the lowest rotational transition line of LiH at $\nu_0 = 444$ GHz to be present. In this case it is convenient to perform narrow-band measurements, since the rotational velocity of dark clusters causes a Doppler shift $\Delta\nu/\nu_0 \sim 3 \times 10^{-4}$ and the line broadening (mainly due to the turbulent velocity $\sim 10$ km s$^{-1}$ of molecular clouds in dark clusters) is $\sim 5 \times 10^{-5}$.

On the other hand, if many lines are present in the microwave band it could be more convenient to perform broad-band measurements which, although dilute the effect, have a higher sensitivity. In the latter case, the Doppler shift effect will show up as an anisotropy in the CBR. This phenomenon is the Doppler shift for a free-streaming gas with speed $v$ (Sunyaev & Zeldovich 1980). The corresponding anisotropy is

$$\frac{\Delta T}{T_r} = \pm \frac{v}{c} \, S \, f \, \overline{\tau}_\nu \, . \tag{1}$$

The sign depends on the direction of the projected velocity along the line of sight and $T_r$ is the temperature of the CBR. The optical depth $\tau_\nu$ depends on the molecular abundance, on the number of molecules in each quantum level and on the resonance cross-section (for details see Melchiorri & Melchiorri 1994). Since the clouds are optically thick only at some frequencies, it is convenient to use in Eq. (1), instead of $\tau_\nu$, an *averaged optical depth* over the frequency range ($\nu_1 \leq \nu \leq \nu_2$) of the detector:

$$\overline{\tau} = (\nu_2 - \nu_1)^{-1} \int_{\nu_1}^{\nu_2} \tau_\nu \, d\nu \, . \tag{2}$$

So, if we could look for anisotropies in the CBR towards M31 with resolution of $\sim 3''$, Eq. (1) would imply anisotropies of $\sim 3 \times 10^{-4} \, f \, \overline{\tau}$ in $\Delta T/T_r$ (since $S \sim 1$).

As it is difficult to work with fields of view of a few arcsec, we calculate below the expected CBR anisotropy between two fields of view (on opposite sides of M31) separated by $\sim 4°$ and with angular resolution of $\sim 1°$. Supposing that the halo of M31 consists of $\sim 10^6$ dark clusters and that all of them lie between 25 kpc and 35 kpc, we would be able to detect $10^3 - 10^4$ dark clusters per degree square. Scanning an annulus of $1°$ width and internal angular diameter $4°$, centered at M31, in 180 steps of $1°$, we would find anisotropies of $\sim 2 \times 10^{-5} \, f \, \overline{\tau}$ in $\Delta T/T_r$ (as now $S = 1/25$). In conclusion, since the theory does not permit to establish whether the expected anisotropy lies above or below current detectability ($\sim 10^{-6}$), only observations can resolve this issue.

However, we should point out that broad-band measurements can be very difficult due to the contribution of cirruses emission in the microwave band. Cirruses in M31 itself should not interfere with the above proposed measurements of the CBR anisotropies since they are inside $\sim 15$ kpc, as results from IRAS observations (Meurs & Harmon 1988). Another problem are foreground cirruses within our galaxy located in the direction towards M31, that IRAS showed to be inhomogeneously distributed (Walterbos & Schwering 1987). Their contribution could be subtracted using the available IRAS maps. On the other hand it is well possible that there exists cold dust whose emission results in a spectrum that peaks beyond the longest wavelength IRAS band ($100 \mu$m) and whose presence could be inferred performing the above proposed measurements in the microwave band with different decreasing angular widths rather than using a fixed width of $1°$ as mentioned. In this case, since the filling factor ($S = 1$) does not change, the presence of cold dust along the field of view manifests itself giving constant values for the CBR anisotropy. Otherwise, in the absence of foreground cold dust, we should observe an increase in $\Delta T/T_r$ values.

## 3. Direct detection of MHOs in M31 via their infrared emission

We assume, as illustration, that all MHOs have equal mass $m \sim 0.08~M_\odot$ (which is the upper mass limit for brown dwarfs), the same age $t \sim 10^{10}$ yr, an effective surface temperature $\sim 1.4 \times 10^3$ K (Adams and Walker 1990, Daly and McLaughlin 1992) and make up the fraction $(1 - f)$ of the dark matter in M31. As a consequence, MHOs emit most of their radiation at the wavelength $\lambda_{max} \sim 2.6~\mu$m ($\nu_{max} \sim 11.5 \times 10^{13}$ Hz). We shall assume that the M31 dark halo (i) is made of uniformly distributed MHOs, (ii) is spherically symmetric and (iii) its mean density goes like $\rho(r) = \rho(0)[1+(r/a)^2]^{-1}$, where now $r$ denotes the distance of a MHO from the M31 centre, $\rho(0) \sim 1.9 \times 10^{-24}$ g cm$^{-3}$ is the central dark matter density while $a \sim 5$ kpc is the dark matter core radius. The surface brightness $I_\nu(b)$ as a function of the projected separation $b$ (impact parameter) of a point in the halo from the M31 centre is given by (Adams and Walker 1990)

$$I_\nu(b) = \frac{1-f}{4\pi}~\Gamma_\nu~\int_{s_{min}}^{s_{max}} ds~\rho(r(s))~, \tag{3}$$

where $\Gamma_\nu$ is the M31 dark halo specific luminosity and $s$ denotes the distance of a generic M31 dark halo point from Earth along the line of sight. Manifestly, $s_{min}$ and $s_{max}$ are the nearest and the farthest intercepts of the M31 dark halo along the line of sight, respectively. The calculation of the integral in Eq. (3) is straightforward and proceeds as in (Jetzer 1994). Moreover, using the explicit expression of $I_\nu$ and $x = \nu/3.8 \times 10^{-1}$ Hz, we get

$$I_\nu(b) \sim 5 \times 10^5 \frac{x^3}{e^x - 1} \frac{a^2(1-f)}{D\sqrt{a^2 + b^2}} \arctan \sqrt{\frac{L^2 - b^2}{a^2 + b^2}} \text{ Jy sr}^{-1}, \qquad (4)$$

where the halo radius is $L \sim 50$ kpc. Some numerical values of $I_{\nu_{max}}(b)$ with $b = 20$ kpc and 40 kpc are $\sim 1.6 \times 10^3$ $(1-f)$ Jy sr$^{-1}$ and $\sim 0.4 \times 10^3$ $(1-f)$ Jy sr$^{-1}$, respectively. For comparison, we recall that the halo of our galaxy would have in the direction of the galactic pole a surface brightness $I_\nu \sim 2 \times 10^3$ Jy sr$^{-1}$ at 3.8 $\mu$m (Adams and Walker 1990). In order to disregard from Eq. (4) contributions from ordinary stars in the M31 disc, the parameter $b$ should be larger than $\sim 10$ kpc, while the signal from the M31 halo can be identified and separated from the background via its b-modulation. Moreover, the infrared radiation originating from the MHOs in the halo of our galaxy can be recognized (and subtracted) by its characteristic angular modulation.

So far, we have been concerned with the infrared detection of uniformly distributed MHOs in M31. What about dark clusters? The typical angular size of a dark cluster with radius $\sim 10$ pc in M31 is $\sim 3''$, but the corresponding intensity of $\sim 4.5 \times 10^{-4}$ mJy is too small for present experimental detection. Hence, it will be impossible to distinguish an overall cluster-like MHO distribution from a uniform one.

Finally, we point out that the angular size of dark clusters in the halo of our galaxy at a distance of $\sim 20$ kpc is $\sim 1.8'$. In addition, the typical separation between two MHO clusters is $\sim 14'$ if their total number is $\sim 10^6$. As a result, a typical pattern of bright and dark spots should be seen by pointing the detector into different directions.

*Acknowledgements.* PhJ would like to thank the Institute for Nuclear and Particle Astrophysics at LBL and the Center for Particle Astrophysics at University of California, Berkeley for hospitality while this work was carried out and Y. Rephaeli for a useful discussion. FDP and GI thank A. Perrone and F. Strafella for instructive discussions. Finally we thank J. Lequeux for many suggestions.